\documentclass{PoS}

\newcommand{\bp}{{\bf p}}
\newcommand{\bq}{{\bf q}}

\newcommand{\bK}{{\bf K}}

\newcommand{\bb}{{\bf b}}
\newcommand{\br}{{\bf r}}
\newcommand{\bC}{{\bf C}}

\newcommand{\bkappa}{\mbox{\bf $\kappa$}}
\newcommand{\ket}[1]{| {#1} \rangle}
\newcommand{\bra}[1]{\langle {#1} |}

\newcommand{\half}{{1\over 2}}
\newcommand{\openone}{I\!\!I}
\def\CutPom{{\bf I\!P\!\!\!\!\!\Big{ / }}}
\def\Pom{{\bf I\!P}}

\title{Nonlinear $k_t$ factorisation: recent progress and perspectives}

\ShortTitle{Nonlinear $k_t$ factorisation: recent progress and perspectives}

\author{\speaker{Wolfgang Sch\"afer}\\%
        Institute of Nuclear Physics PAN, ul. Radzikowskiego 152, 31-342
        Krakow, Poland \\
        E-mail: \email{wo.schaefer@fz-juelich.de}}

\abstract{Hard scattering in a strongly absorptive regime requires 
a novel nonlinear $k_\perp$-- factorization.
Here we discuss two recent developments: firstly the evaluation
of radiative corrections to single particle spectra, and secondly an
extension of the formalism to address topological cross sections and 
unitarity cutting rules. 
}

\FullConference{Diffraction 06, International Workshop on Diffraction in High-Energy Physics\\
                 September 5-10, 2006\\
                 Adamantas, Milos island, Greece}

\begin{document}

\section{Introduction}

Within perturbative QCD (pQCD) at high energies it is by now well understood that
an adequate treatment should account explicitly for the
transverse--momentum 
degrees of freedom of the participating partons. 
Specifically in the high--energy limit, this is accomplished by the 
so-called $k_\perp$-factorization which  ultimately goes back to  
works on the Balitsky-Fadin-Kuraev-Lipatov (BFKL) equation for the
small-$x$ evolution of the unintegrated glue (\cite{BFKL}, for
recent reviews on $k_\perp$-factorization and more references see
\cite{Smallx,Antoni}). An underlying assumption of the phenomenology 
based on (linear--) $k_\perp$--factorization is that it is sufficient to
consider only single gluon (reggeon) exchange in the parton--target
interaction. Certainly, this can only be warranted in a finite energy 
range, although perhaps covering most of the small--$x$ domain 
of $\gamma^* p$ collisions at HERA. Indeed, in a color--dipole picture
\cite{NZ91}, HERA data on diffractive deep--inelastic scattering imply,
that for color dipoles $\sigma_{el}/\sigma_{tot} \ll 0.5$, that is unitarity constraints 
must be weak. Still, eventually the $s$--channel partial waves 
of single--gluon exchange will overshoot unitarity bounds,
and the inclusion of multiple gluon/reggeon exchanges associated with 
absorptive/unitarity corrections is mandatory. Although some 
semi--quantitative phenomenology exists (e.g.\cite{Barone}), 
a systematic theory for proton targets is absent. Here heavy nuclei as targets 
offer a welcome testing ground for unitarity effects. 
Multiple gluon exchanges are enhanced by the large thickness of a heavy
target nucleus, and one can concentrate on those unitarity corrections 
which grow with the size of the target $A$. 
From a different viewpoint \cite{Mueller}, the opacity of heavy nuclei generates a new
large scale $Q_A^2 \propto T_A(\bb)$, the saturation scale, where $T_A(\bb)$
is the nuclear thickness (see e.g. \cite{KovchegovReview} for a review
and references). Especially in view of present 
and future programs on forward physics at RHIC and LHC,
the fate of $k_\perp$ factorization in a regime of strong nuclear
absorption is an important issue. 

For illustration, consider the production of dijets in the virtual
photon fragmentation region in $\gamma^*$--proton collisions.
Here the dominant contribution comes from the $q\bar{q}$
final states and can be viewed as a hard photon--gluon fusion 
$ \gamma^* g_t \to q\bar{q}$. While in the familiar collinear 
factorization the final state is a pair of back--to--back jets, in the
lowest order of $k_\perp$--factorization the quark and antiquark 
jets are decorrelated in azimuth, and the 
distribution in the decorrelation momentum maps out the 
proton's unintegrated glue \cite{HERADijets}:
\begin{equation} 
{d \sigma (\gamma^* \to q \bar{q}) \over dz d^2\bK d^2\bp} \propto  f(x,\bK) \Big|\psi(z,\bp) -
\psi(z,\bp - \bK) \Big|^2 
\label{eq:linear} 
\end{equation}
Here $\bp$ is the transverse momentum of the quark--jet, $z$ is the
fraction of the photon's lightcone momentum carried by the quark, and
the transverse momentum $\bK$ transferred by the gluon quantifies the
azimuthal decorrelation of the $q\bar{q}$--system. Finally $\psi(z,
\bp)$ is the momentum space lightcone-wavefunction (LCWF) for
the transition $\gamma^* \to q\bar{q}$, and $f(x,\bK) \propto \bK^{-4}
\partial G(x,\bK) / \partial \log (\bK^2) $ denotes the unintegrated
gluon distribution of the proton target, which e.g. is directly
proportional to the amplitude of hard diffractive dijet production
\cite{NZSplitting}.
When addressing the same observable in deep inelastic scattering (DIS)
off nuclear targets \cite{Nonlinear}, using a nuclear unintegrated glue that fulfills the
same proportionality to hard diffractive amplitudes \cite{NSS}, we found
that the linear $k_\perp$--factorization is completely broken. The
spectrum of dijets off nuclei is an involved nonlinear functional with
little resemblance\footnote{See eq.(\ref{eq:TopXS}) below.}  of the free--nucleon result eq. (\ref{eq:linear}). It
incorporates distinct physical effects, such as the {\it{coherent}}
nuclear distortion of the $q\bar{q}$ LC-WF as well as {\it{incoherent}}
transverse--momentum broadening through intranuclear multiple
scattering, which arise from the same unitarity--driven dynamics.
We dubbed the emerging momentum--space formalism {\it Nonlinear $k_\perp$
  Factorization}, and subsequently developed the description of other 
relevant hard subprocesses, such as $\pi \to q\bar{q}, \, q \to qg,\, g \to gg  \, , g \to
q\bar{q}$
\cite{PionDijet,SingleJet,Nonuniversality}.
In this contribution we will address the most recent
developments: first, the evaluation of virtual radiative corrections to
single--jet observables, and the small--$x$ evolution of the nuclear
unintegrated glue from the $s$--channel unitarity relations
\cite{RealVirtual}, and second an extension to more refined observables, such as topological
cross sections and the related unitarity rules \cite{Topological} of 
Abramovsky--Gribov--Kancheli (AGK \cite{AGK}) type. 
This will include prolegomena to a Reggeon Field Theory (RFT) 
interpretation of the nonlinear $k_\perp$--factorization results.
The breaking of linear $k_\perp$--factorization
has also been reported in other approaches, we mention
\cite{Blaizot}. 

\section{Nuclear unintegrated glue \cite{Nonlinear,NSS}, and its
  evolution \cite{RealVirtual}}

Swiftly moving partons propagate along fixed impact parameter
trajectories, and their impact parameter is conserved in the interaction
with the target $A$. Consider a parton $a$ in an arbitrary color representation, and
let $\textsf{S}_a(\bb_a) $ be the $S$--matrix of the
(bare) parton--target interaction in the target rest frame at an
impact parameter $\bb_a$. Then, the $S$--matrix for the $a \bar{a}$ 
color--dipole of size  $\br = \bb_a - \bb_{\bar{a}}$ 
scattering at $\bb$ 
\begin{equation}
\textsf{S}_{a\bar{a}}(\bb,\br) =
{\bra{A}{\rm Tr}[\textsf{S}_a(\bb_a)\textsf{S}_a^\dagger(\bb_a-\br)] 
\ket{A} \over \bra{A} {\rm Tr} \openone  \ket{A} }
\nonumber\\
\end{equation}
can be evaluated by standard Glauber--Gribov\cite{Glauber,Gribov} techniques at a boundary
value $x_A = (2R_Am_N)^{-1}$ ($R_A$ is the nuclear radius, $m_N$ the
nucleon mass) as $\textsf{S}_{a\bar{a}}(\bb,\br) = \exp[-\half
\sigma_{a\bar{a}}(x_A,\br) T_A(\bb) ]$, 
and serves to define the nuclear unintegrated gluon distribution $ \phi(\bb,x_A,\bp)$
\begin{equation}
\Phi(\bb,x_A,\bp) = \int {d^2 \br \over (2 \pi)^2}
  \textsf{S}_{a\bar{a}}(\bb,\br) \exp[-i \bp \br] =
  \exp[-\nu_A(\bb)] \delta^{(2)}(\bp) + \phi(\bb,x_A,\bp) \, .
\label{eq:unintegrated}
\end{equation}
Using its Glauber--Gribov relation to the free--nucleon color--dipole
cross section $\sigma(x_A,\br)_{a\bar{a}}$ one derives a convenient
representation of the coherent nuclear glue $\phi(\bb,x_A,\bp)$ in terms
of multiple convolutions $f^{(j)}(\bp) = (f \otimes f^{(j-1)}) (\bp) $ of
  the free--nucleon unintegrated glue $f(\bp)$: $  \phi(\bb,x_A,\bp) =
\sum w_j(\nu_A(\bb))  f^{(j)}(\bp) $. 
Here the Poisson weights 
\footnote{The nuclear opacity
    $\nu_A(\bb) = \half \sigma_0(x_A) T_A(\bb)$ involves the dipole cross
    section for large color dipoles $\sigma_0(x_A)$ and the nuclear
    thickness function $T_A(\bb)$.} 
$w_j
=\exp[-\nu_A(\bb)] \nu_A^j(\bb)/j!$ 
represent the probability to find 
$j$ overlapping nucleons at impact parameter $\bb$. The salient features
of $\phi(\bb,x_A,\bp)$ can be briefly summarized: first, for realistic
free--nucleon input and heavy nuclei there emerges a saturation scale
$Q^2_A(x_A) \sim 0.8 \div 1$  GeV$^2$, second, at large $\bp^2$ we have
a Cronin--type antishadowing enhancement that is calculable
parameter--free, third it furnishes a {\it{linear}}
$k_\perp$--factorization of inclusive DIS, forward single--jets in DIS
and diffractive dijet amplitudes. To study evolution to small $x \ll
x_A$, one has to follow the strategy of \cite{NZ94}, and integrate
out the effect of an additional s-channel gluon, schematically:
\begin{equation}
\delta  \textsf{S}_{a\bar{a}}(x,\bb,\br) \propto \log \Big({x_A\over x}\Big)
\int |\psi_{a\bar{a} g}|^2 \Big(
\textsf{S}_{a\bar{a}g} -  \textsf{S}_{a\bar{a}} \Big ) \, .
\label{eq:evolution}
\end{equation}
This produces the first step of the small--$x$ evolution, a closed
form of which does not exist, and is equivalent to the first step of 
the Balitsky--Kovchegov (BK) equation \cite{BK}. Using the relation (\ref{eq:unintegrated})
one may translate (\ref{eq:evolution}) to momentum space. Concentrating
on color-triplet partons (quarks), one would obtain \cite{RealVirtual}
\begin{equation}
\delta \phi(\bb,x,\bp) = \log\Big({x_A \over x}\Big) \Big[ ({\cal{K}}_{BFKL} \otimes
\phi)(\bb,x_A,\bp) + {\cal{Q}}[\phi](\bb,x_A,\bp) \Big] \, . 
\label{eq:evolution}
\end{equation}
Here we decomposed 
the nuclear evolution into the linear, BFKL-evolution piece, $({\cal{K}}_{BFKL} \otimes
\phi)(\bb,x_A,\bp) = C_A \alpha_S (2 \pi^2)^{-2} \int d^2\bq [ 2
K(\bp,\bp-\bq)  \phi (\bb,x_A,\bq)  - \phi (\bb,x_A,\bp) K(\bq,\bq-\bp)
]$, and the genuinely nonlinear component represented by the quadratic
functional \footnote{here, for perturbative large $\bp_i$, $ K( \bp_1, \bp_2 ) = (\bp_1 - \bp_2 )^2 /
  \bp_1^2 \bp_2^2$. For a regularized form valid also at small $\bp_i$,
see \cite{RealVirtual}.} 
\begin{eqnarray} 
{\cal{Q}}[\phi](\bb,x_A,\bp)
&&=  \int d^2\bq_1 d^2\bq_2 \phi(\bb,x_A,\bq_1) \{ [K( \bp + \bq_1, \bp +
\bq_2) - K( \bp , \bp + \bq_1 ) \nonumber \\
&&- K(\bp, \bp + \bq_2) ] \phi(
\bb,x_A,\bq_2) \nonumber 
- \phi(\bb,x_A,\bp) [ K ( \bq_1, \bq_1 + \bq_2 + \bp) - K
( \bq_1 , \bq_1 + \bp) ] \} \, . 
\end{eqnarray}
The latter generates a higher--twist
contribution to the large--$\bp$ tail of the nuclear collective glue,
but it does not exhaust the nuclear higher twists, recall that the boundary
condition itself contains substantial -- but {\it{anti}}-shadowing -- higher
twist contributions. 
 
\section{Forward jets from the breakup of valence quarks \cite{RealVirtual}}

\begin{figure}[!t]
\begin{center}
\includegraphics[width = .3\textwidth,angle=270]{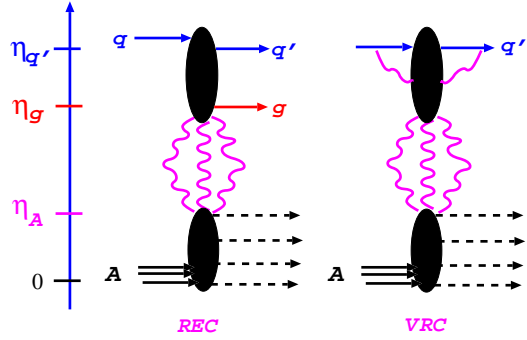}
\caption{The real emission (REC) and virtual radiation (VRC) contributions
to the spectrum of leading quark jets in inclusive production off
nuclei. The (pseudo)rapidity of radiation correction gluons runs
between the quark jet rapidity and the nuclear boundary condition
rapidity $\eta_A=\log(1 / x_A)$, where $x_A$ is defined in the text.}
\label{fig1}
\end{center}
\end{figure}

Consider now the production of jets/high $p_\perp$ particles at forward
rapidities, i.e. in the beam fragmentation region  
in a high energy proton--nucleus collision. The dominant beam partons
are evidently valence quarks. The density of valence quarks is a 
steep function for $x \to 1$, and the production of high $p_\perp$
forward partons must be highly sensitive to their energy loss.
Here we discuss the perturbative, radiative energy loss that originates
from the transition $q \to qg$. The dynamics of partons with rapidities $\eta > \eta_A =
\log(1/x_A) $, proceeds coherently over the
whole nucleus (for some notation see Fig.~\ref{fig1}). Notice that in addition
to the coherent excitation $q \to qg$, in the $k_\perp$ factorization
one must also add the quasielastic scattering of the beam quark, where 
its transverse momentum is balanced by (mini--) jets with rapidities 
$\eta < \eta_A$. To quantify the $p_\perp$--dependent radiative energy
loss, one must equally evaluate the radiative correction to the
quasielastic scattering, i.e. the virtual corrections in terms of fast
$\eta > \eta_A$ parton degrees of freedom depicted as VRC in Fig.~\ref{fig1}.
The calculation of radiative corrections, real as well as virtual is 
most expediently performed in impact parameter space. To this end,
one expands the LCWF of the beam quark into its bare quark and
quark--gluon components. By the conservation of impact parameters, the
$S$--matrix acts on the physical quark state in a simple form:
\begin{equation}
\textsf{S}\ket{q_{phys}(\bb)} = [S_q(\bb) + \delta S_q(\bb)] \ket{q_{phys}(\bb)} +
 [ S_q(\bb_q') S_g(\bb_g) - S_q(\bb) ] \psi_{qg}(z_b,\bb_q'- \bb_g) \ket{q(\bb_q') g(\bb_g)} \, . 
\end{equation}
Evaluation of two--parton and single--parton spectra from the excitation
operator $S_{ex} =  S_q(\bb_q') S_g(\bb_g) - S_q(\bb) $ is discussed in
detail
in \cite{SingleJet,Nonuniversality}. The important point
is that $S_{ex}$ determines through the unitarity relation also the radiative
correction $ \delta S_q(\bb)$ to the quasielastic scattering of the
quark, which reads:
\begin{equation}
\delta S_q(\bb) = \int {\cal{D}}\{\bb_g,z_g\} |\psi_{qg}|^2 [S_q(\bb_q')
S_g(\bb_g) - S_q(\bb)] \, . 
\end{equation}
It is important to stress, that the LCWF/color--dipole S--matrix
approach is fully general in that it applies equally to the
single--gluon exchange encountered on the free nucleon as well 
as to the multigluon exchanges relevant to strongly absorbing nuclei.
Also notice, that we {\it{do not require}} the multi--Regge type kinematics 
of $\eta_g \ll \eta_q$, but allow for an arbitrary rapidity spacing of
scattered quark and the radiative gluon as is appropriate for the
evaluation of the radiative energy loss. We skip the detailed form of
the single--parton spectra, as well as the case of incident color--octet
partons, and close this section with a remark on the small--$x$
evolution properties of the single--quark spectra.  
At the boundary value $x_A$ one can easily check that the latter is
directly proportional to the target unintegrated glue, 
$ d \sigma_{Qel} ( q \to q , x_A) / d^2\bb d^2 \bp = \phi(\bb,x_A,\bp)$.
The intriguing result is now that this relation holds, to the
leading--log$(1/x)$ (LL$(1/x)$) also {\it{after}} the radiative corrections,
that is
\begin{equation}
{d \sigma_{Qel} ( q \to q , x) \over d^2\bb d^2 \bp} = \phi(\bb,x_A,\bp)
+\delta\phi(\bb,x,\bp)  \, ,
\end{equation}
with {\it{the same}} $\delta\phi(\bb,x,\bp)$ of eq.(\ref{eq:evolution}) 
that defines the first
LL$(1/x)$
iteration of the nonlinear small--$x$ evolution of the nuclear
unintegrated glue. We recall also the earlier result \cite{SingleJet}, 
that the spectrum of midrapidity gluons is linearly factorizable in terms of the nuclear
unintegrated glue. 
All of this suggests a deeper role of the coherent nuclear glue in terms
of a to be developed RFT description. Some of these aspects we shall
address in the following.

 \section{Topological cross sections and unitarity cutting rules \cite{Topological}}

In linear $k_\perp$ factorization the partonic transitions $a \to bc$ of
interest proceed via single gluon exchange and leave the recoiling
target nucleon debris in a color octect state. There is an obvious 
connection between the unintegrated gluon distribution and a cut (BFKL-)
Pomeron. For example, the quasi--elastic scattering of parton $a$ with
a color excited nucleon in the final state is $d \sigma_{Qel}/d^2\bp =
C_a/2C_F f(x,\bp)$, where $C_a,C_F$ are color Casimirs and the
nuclear unintegrated glue $f$ represents the coupling of the
color--octet $t$--channel exchange to the nucleon.
The situation is more involved for the heavy nucleus target.
On the {\it{beam side}}, during its coherent interaction with the
nucleus the $bc$ parton system will evolve over all possible 
color multiplets in the relevant product representation space. 
The at first sight formidable problem in color algebra can be solved
\cite{Nonuniversality}
using a technique originally due to Zakharov, where by concentrating
on the $bc\bar{b}'\bar{c}'$--density matrix one deals with a
four--parton system in an overall color--singlet state.
The intranuclear evolution problem is then solved by a 
color-coupled channel four--body $S$--matrix $S(\bC) = \exp [-\half
\Sigma^{(4)}(\bC) T_A] $ (here $\bC$ collectively denotes the relevant
impact parameters). The building block is the color--dipole cross
section operator $\Sigma^{(4)}(\bC)$ for the $bc\bar{b}'\bar{c}'$
system for the free--nucleon target. It is a matrix in the space of
possible color-singlet states $\ket{R\bar{R}} = \ket{(bc)_R \otimes
  (\bar{b}' \bar{c'})_{\bar{R}}}$. The overall--singlet property 
of the $bc\bar{b}'\bar{c}'$--system implies that its matrix elements are
infrared safe combinations of the free--nucleon color dipole cross
section. It is important to distinguish the following aspects of color
channel coupling: {\bf{a}}) transitions of the $bc$--system between {\it{two
multiplets, $R_i \to R_j$ of different dimensionality}}. While these transitions
are in general suppressed in the number of colors $N_c$, a large number
of final states in higher multiplets can overcome this suppression. The
$t$--channel gluons that induce these transitions must be {\it{hard}} 
to resolve the color structure. Correspondingly, such color excitations 
can be treated perturbatively in an expansion, that is simultaneously 
a large--$N_c$/hard scattering expansion. {\bf{b}}) transitions of the
$bc$--systems that are {\it{rotations within the same color
multiplet}} $R_i \to R_i$. Notice that these can involve singlet, as well as octet
exchange with a scattering center in the target. Such contributions need
to be summed to all orders and place no restriction on the hardness of
the $t$--channel gluons. 

\begin{figure}[!t]
\begin{center}
\includegraphics[width = .5\textwidth]{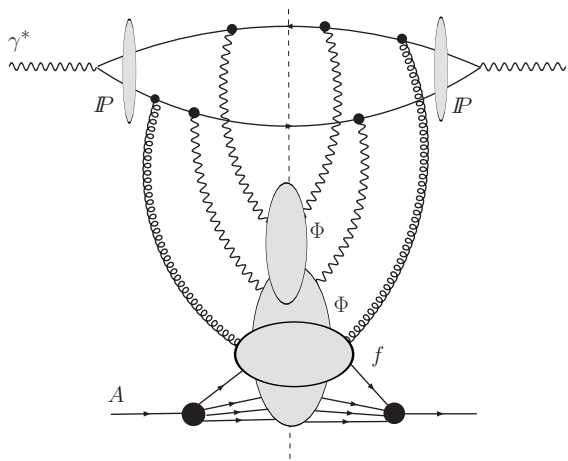}
\caption{One of four diagrams representing the RFT unitarity cut for
  inelastic DIS in terms of two types of cut Pomerons. The curly lines
represent gluons from the cut Pomeron
 $\CutPom_e$ associated with the
 singlet to octet transition of the $q\bar{q}$ pair, and couples to the
 target through the free--nucleon glue $f$. Zigzag-lines are 
 associated with two cut nuclear Pomerons $\CutPom_{r,A}$ associated with
 color rotations in the $q\bar{q}$--octet channel. Their coupling through
 the target is through the nuclear unintegrated glue $\Phi$. 
 When expanded out they give rise to the factors $\propto
 d\sigma_{Qel}^{(k)}(\bkappa_1) 
 \cdot d\sigma_{Qel}^{(j)}(\bkappa_2)$ 
 in eq. (\protect \ref{eq:TopXS} ). Also indicated are the coherent distortions of the $\gamma^* \to q\bar{q}$
LCWF through uncut Pomerons. Notice the cylinder/Mandelstam structure 
of the cut. The coupling of multiple Pomerons is local in rapidity and
involves no branching tree of triple Pomeron interactions/fan diagrams. }
\label{fig:RFT}
\end{center}
\end{figure}

When viewed from the {\it{target side}},
after the scattering the nucleus will be left in a state with multiple
color excited nucleons 
\footnote{Note a certain similarity to the concept of wounded nucleons \cite{Bialas}.}
.
The higher the number of color excited nucleons,
the higher will also be the multiplicity of particles produced in the 
target hemisphere. This clearly implies nonperturbative mechanisms
for energy loss of the $bc$ system. In addition, this picture offers a perfect
definition of cut Pomerons and connection with the RFT language: 
each and every color excited nucleon must be associated with a 
unitarity cut of a Pomeron exchanged between beam and target.
From a technical point of view, the counting of color--octet nucleons
is possible after a decomposition of the dipole cross section--operator 
$\Sigma^{(4)}(\bC) = \Sigma_{el}^{(4)}(\bC) + \Sigma^{(4)}_{ex}(\bC)$ 
into an 'elastic' piece $\Sigma^{(4)}_{el}$ that involves two--gluon color
singlet exchange with the nucleon, and an excitation piece
$\Sigma^{(4)}_{ex}$ that involves color-octet exchange with a nucleon
\footnote{Such decomposition is clearly infrared sensitive, but so is
also the transition from cut Pomerons to final state particle
multiplicities.}
.
We come to the implications on the unitarity cut interpretation of the
nuclear unintegrated glue. Firstly, one would easily identify the cross 
section for $j$--fold quasielastic scattering of a parton $a$ as
$d\sigma^{(j)}_{Qel}/d^2\bp = C_a/2C_F f^{(j)}(\bp)$. The familiar
expansion of the nuclear unintegrated glue over multiple convolutions
now takes a new interpretation as an expansion over unitarity cuts 
corresponding to final states with $j$ color excited nucleons: 
$\phi(\bb,x,\bp) = \sum w_j(\nu_A(\bb) )
d\sigma^{(j)}_{Qel}/\sigma_{Qel}d^2\bp $. In the alternating sign
expansion $\exp[-\nu_A(\bb)] = \sum (-1)^k/k! [T_A/2]^k [\int d^2\bp
f(\bp)]^k $ of the exponential entering $w_j$, the $k$-th order term
corresponds precisely to an absorptive correction from $k$ uncut 
Pomeron exchanges. This suggests an interpretation of $\phi(\bb,x,\bp)$ 
as a coupling of the absorbed gluon--Reggeon to the nucleus.
Out of the many possible applications we give a brief description 
on the RFT-cut interpretation of forward--dijets in inelastic DIS --
the counterpart of  the free nucleon result eq.(\ref{eq:linear}).
We obtain the nonlinear $k_{\perp}$-factorization for 
topological cross sections of DIS followed by color excitation of 
$\nu$ nucleons ($\nu$ cut pomerons) 
   \footnote{Here $\beta$ is a
  dimensionless longitudinal depth of the nucleus, which arises because
  of the noncommutativity of $\Sigma_{el}^{(4)}$ and
  $\Sigma^{(4)}_{ex}$, and
  $\psi(\beta,...)$ denotes a LCWF 
   coherently distorted over a slice of length $\beta$.}: 
\begin{eqnarray}
&&{d\sigma_{\nu} (\gamma^*A\to \{q\bar{q}\}_8  X)\over d^2\bb dz d^2\bp
  d^2\bK} =
{T_A(\bb)\over (2\pi)^2} \int_0^1 d\beta 
\int  d^2\bkappa_1 d^2\bkappa_2 d^2\bkappa 
\delta(\bK -\bkappa_1-\bkappa_2-\bkappa)
\nonumber\\
&\times& {d\sigma_{Qel}(\bkappa) \over d^2\bkappa}
\Big|\psi(\beta;z,\bp-\bkappa_1)
-\psi(\beta;z,\bp-\bkappa_1 -\bkappa)\Big|^2
\nonumber\\ 
&\times& \sum_{j,k=0} \delta(\nu-1-j-k) 
w_j\Big((1-\beta)\nu_A(\bb)\Big)
w_k\Big((1-\beta)\nu_A(\bb)\Big)
{d\sigma_{Qel}^{(k)}(\bkappa_1) \over \sigma_{Qel} d^2\bkappa_1}
\cdot{d\sigma_{Qel}^{(j)}(\bkappa_2)
\over \sigma_{Qel} d^2\bkappa_2}.
\label{eq:TopXS}
\end{eqnarray}
The color channel coupling effects manifest themselves in the presence 
of two different types of cut Pomerons, here they show up as
$d\sigma_{Qel}(\bkappa) / d^2\bkappa \propto f(\bkappa)$, which derives from the
singlet--to--octet transition of the $q\bar{q}$--pair, whereas 
$d\sigma_{Qel}^{(k)}(\bkappa_1) / \sigma_{Qel} d^2\bkappa_1
\cdot d\sigma_{Qel}^{(j)}(\bkappa_2)
/ \sigma_{Qel} d^2\bkappa_2 $ derive from color rotations of the $q
\bar{q}$--pair in the color--octet state; quark and antiquark rescatter
independently and contribute $j$, respectively $k$ color excited
nucleons to the final state.  The RFT structure is further explained in
Fig.~\ref{fig:RFT}.  When going over to topological cross sections for
single--jet production, say, the single quark spectrum, interactions
of the spectator anti--quark do not cancel but leave a Cheshire Cat
Grin (CCG): the final state will contain color excited nucleons that derive
from interactions of the antiquark. Amazingly, a novel multiplicity
resummation, removes the CCG. The contribution from $k$ cut Pomerons
can be isolated from 
\begin{equation}
{d\sigma^{(k)} (\gamma^*A\to q X)\over d^2\bb dz d^2\bp}= 
\sum_{\nu> k} {d\sigma_{\nu} (\gamma^*A\to q X)\over d^2\bb dz d^2\bp}
\, ,
\end{equation}
which roughly corresponds to a re--summation over backward
multiplicities $n_B > k \langle n_B\rangle_{pp} $.

As a consequence of {\it{two types of cut Pomerons}}, common doctrine
from hadronic Glauber--type models \cite{GlauberAGK}
is entirely inapplicable in the pQCD realm. For example an expansion of the inelastic
Glauber profile function for color dipoles 
\begin{equation}
\Gamma^{(inel)} = 1 -
\exp[-\sigma(x,\br) T_A(\bb) ] = \exp[-\sigma(x,\br) T_A(\bb)] \sum
{1 \over \nu !} [\sigma(x,\br) T_A(\bb)]^\nu  
\end{equation}  
simply has {\it{no sensible
    unitarity interpretation at all}}. In particular it does not
allow one to extract topological cross sections from the inclusive
inelastic profile function. Additionally, relations on the
double Pomeron contributions that are in widespread use as a
unitarisation snake--oil, such as the cancellation of discontinuities
\begin{equation}
\Delta_2 \Gamma^{(in)} (\CutPom \Pom) + 2  \Delta_2 \Gamma^{(in)} (\CutPom \CutPom)
= 0
\end{equation}
and the ``AGK-ratios'' 
\begin{equation}
\Delta_2 \Gamma_{D}(\Pom\Pom) : \Delta_2 \Gamma_{1}^{in}(\CutPom \Pom) :
    \Delta\Gamma_{2}^{in}(\CutPom\CutPom)= 1:-4:2 \, , 
\end{equation}
take no account of the intricacies of color channel coupling (two types
of cut Pomerons) and the subtleties of spectator interactions. See
\cite{Topological} for the --infrared sensitive-- relations that replace
them. Gribov's unitarity relation $\Delta_2 \Gamma_{tot}(\Pom \Pom) =
- \Delta_2 \Gamma_{Diffractive} (\Pom \Pom)$ between the shadowing
correction to the total cross section and the diffractive profile function is model
independent, and remains of course valid.

\section{Summary}

Hard scattering in a regime of strong opacity  requires a novel,
nonlinear $k_\perp$--factorization. Here we would not reiterate our points
which we made clear enough in the main text, rather we end with an outlook:
there remains a long shopping list of phenomenological applications
connected  to the production of forward high $p_\perp$--particles/jets,
such as: nonperturbative and perturbative quenching, its centrality
dependence, forward--backward correlations 
between hard production in the beam-- and multiplicities in the backward
hemispheres, azimuthal decorrelations, etc, and we look forward to
practical applications. On the theoretical side a study of multipomeron 
couplings and the RFT--interpretation of the small--$x$ evolution 
of the coherent nuclear glue are to be ordered for Diffraction2008.
\\

It is a pleasure to thank the organisers, especially Igor Ivanov and
Alessandro Papa, for their efforts, and for invitation to this exciting workshop.

\end{document}